\documentclass[final,12pt]{clear2026} % Final submission

\title[Variance reduction combining pre-experiment and in-experiment data]{Variance reduction combining pre-experiment and in-experiment data}
\usepackage{times}
\clearauthor{%
 \Name{Zhexiao Lin} \Email{zhexiaolin@berkeley.edu}\\
 \addr Department of Statistics, University of California, Berkeley, CA 94720, USA
 \AND
 \Name{Pablo Crespo} \Email{pcrespo@etsy.com}\\
 \addr Etsy, Inc., Brooklyn, NY 11201, USA%
}

\usepackage{amsfonts}
\usepackage{tikz}
\usetikzlibrary{arrows,matrix,positioning,calc,automata,patterns}

\def\given{\,|\,}
\def\bR{{\mathbb R}}
\def\cS{{\mathcal S}}
\def\E{{\mathrm E}}
\def\P{{\mathrm P}}
\def\Var{{\rm Var}}
\def\Cov{{\rm Cov}}
\newcommand{\indep}{\perp \!\!\!\perp}

\begin{document}

\maketitle

\begin{abstract}%
Online controlled experiments (A/B testing) are fundamental to data-driven decision-making in many companies. Improving the sensitivity of these experiments under fixed sample size constraints requires reducing the variance of the average treatment effect (ATE) estimator. Existing variance reduction techniques such as CUPED and CUPAC use pre-experiment data, but their effectiveness depends on how predictive those data are for outcomes measured during the experiment. In-experiment data are often more strongly correlated with the outcome, but using arbitrary post-treatment variables can introduce bias. In this paper, we propose a general, robust, and scalable framework that combines both pre-experiment and in-experiment data to achieve variance reduction. Our framework is simple, interpretable, and computationally efficient, making it practical for real-world deployment. We develop the asymptotic theory of the proposed estimator and provide consistent variance estimators. Empirical results from multiple online experiments conducted at Etsy demonstrate substantial additional variance reduction over current pipeline, even when incorporating only a few post-treatment covariates. These findings underscore the effectiveness of our framework in improving experimental sensitivity and accelerating data-driven decision-making.
\end{abstract}

\begin{keywords}%
average treatment effect, in-experiment data, online controlled experiments, regression adjustment, variance reduction.
\end{keywords}

\section{Introduction}\label{sec:intro}

Online controlled experiments, commonly known as A/B testing, play a pivotal role in data-driven decision-making across numerous industries \citep{kohavi2020trustworthy}. These experiments allow organizations to rigorously evaluate the impact of new features, products, and algorithmic enhancements on key business metrics by systematically comparing different variants with randomized assignment. The core statistical measure derived from these experiments is the Average Treatment Effect (ATE), which quantifies the causal impact of the treatment on the outcome of interest compared to a control. Accurately estimating the ATE facilitates informed product development, optimizes user experience, and increases business revenue. A significant challenge in online controlled experiments is enhancing the sensitivity of the experiments under fixed or constrained sample sizes. Increasing sample sizes to achieve higher sensitivity can be prohibitively expensive or logistically impractical in some cases. Consequently, reducing the variance of the ATE estimator is important, as it allows organizations to detect effects more quickly and accelerates iterative experimentation and product innovation cycles.

A common strategy to achieve variance reduction while preserving estimator consistency involves regression adjustment using auxiliary covariates \citep{freedman2008regression, lin2013agnostic}. In the specific context of online controlled experiments, methods such as CUPED (Controlled-experiment Using Pre-Experiment Data) \citep{deng2013improving, deng2023augmentation} have been developed, leveraging linear regression adjustment with pre-experiment data. Extending this idea, CUPAC (Control Using Predictions as Covariates) \citep{tang2020control} incorporates complex machine learning models for predicting outcomes, enhancing variance reduction capabilities. These methods exploit correlations between pre-experiment auxiliary information and experimental outcomes, improving precision in estimating the ATE. However, the effectiveness of CUPED and CUPAC is inherently constrained by the predictive power of pre-experiment data on the outcome, which, being collected prior to the treatment assignment, may have limited direct relevance to outcomes measured during the experiment.

In contrast, data collected during the experiment itself (in-experiment data) can be much more strongly correlated with the outcome and therefore potentially more useful for variance reduction. The difficulty is causal rather than predictive: a post-treatment variable that lies on a pathway from treatment to outcome is a mediator, and adjusting for it removes part of the treatment effect. Many post-treatment covariates, despite being observed after treatment assignment, remain unaffected by the treatment and are balanced across treatment arms. These variables are different from classical mediators. For instance, in online commerce experiments focused on purchase or payment outcomes, post-treatment covariates such as the number of product detail views, session duration prior to purchase, and add-to-cart actions typically exhibit strong correlations with the final outcomes yet remain unaffected by common UI-level interventions. This independence arises because UI-level A/B tests usually modify specific components like ranking algorithms or button designs without altering broader user navigation patterns or the underlying catalog structure. Similarly, in streaming platforms, post-treatment covariates such as viewing duration, frequency of content interactions, or watchlist updates collected during experiments can closely predict user engagement outcomes, yet remain insensitive to minor interface adjustments or recommendation algorithm changes. When such variables are strongly predictive of the outcome, they can play a role analogous to pre-treatment covariates while offering better contemporaneous signal. Despite their predictive potential, existing variance reduction approaches used in online controlled experiments, such as CUPED and CUPAC, only utilize pre-experiment data, overlooking the informative value in data generated during the experiment. The methodological question is therefore not whether arbitrary post-treatment data should be used, but how to identify the subset that can be used safely.

In this paper, we introduce a novel framework that simultaneously leverages both pre-experiment and in-experiment data to reduce the variance of ATE estimators. Our proposed method leaves the first-stage CUPAC model unchanged and adds a linear second-stage adjustment using selected post-treatment covariates. This two-stage method achieves variance reduction while preserving simplicity, interpretability, and computational efficiency. Importantly, our method avoids reliance on restrictive assumptions such as surrogacy or principal ignorability, which have commonly characterized previous literature on post-treatment covariates. Our ATE estimators are consistent and asymptotically normal, and we provide consistent variance estimators. We illustrate the method on multiple Etsy experiments. Incorporating only a small number of selected post-treatment covariates yields substantial gains beyond the standard CUPAC pipeline.

% The remainder of this paper is organized as follows. In Section~\ref{sec:setup}, we introduce the potential outcomes framework and review existing variance reduction techniques, particularly emphasizing their application in online controlled experiments, including CUPED, CUPAC, and related methods. Section~\ref{sec:method} details the motivation behind our method, formally presents the proposed framework, outlines its theoretical properties, and discusses practical implementation considerations. Section~\ref{sec:exp} provides empirical results from real-world experiments conducted at Etsy, highlighting practical advantages and demonstrating superior performance compared to existing methods. Finally, we conclude in Section~\ref{sec:disc} with broader discussions and potential future research directions.

\vspace{-1.1em}
\section{Problem setting and related work}\label{sec:setup}

\subsection{Potential outcomes framework}

We adopt the potential outcomes framework \citep{rubin1974estimating} and model our problem from a superpopulation perspective. Although we focus on the binary treatment and scalar outcome case, our method and results can be easily generalized to multiple treatment and multivariate outcome cases.

Consider $n$ independent and identically distributed (i.i.d.) samples $\{(W_i, X_i, Y_i(1), Y_i(0))\}_{i=1}^n$ drawn from a superpopulation with joint distribution $(W, X, Y(1), Y(0))$. In practice, we observe the dataset $\{(W_i, X_i, Y_i)\}_{i=1}^n$, where $Y_i = Y_i(W_i)$ is the observed outcome corresponding to the assigned treatment. Here, $W_i \in \{0,1\}$ is the treatment assignment indicator for the $i$-th unit - $W_i = 1$ if the unit is assigned to the treatment group and $W_i = 0$ if assigned to the control group. The variables $Y_i(1) \in \bR$ and $Y_i(0) \in \bR$ represent the potential outcomes under treatment and control, respectively, for unit $i$. We incorporate auxiliary covariates $X_i \in \mathbb{R}^d$, which are used for regression adjustment to improve estimation efficiency. We focus on the case where $X_i$ are pre-treatment covariates. In online controlled experiments, treatment assignments are typically generated through Bernoulli trials with a fixed probability $p \in (0,1)$, ensuring that each unit has an equal and independent chance of receiving the treatment. Formally, we have $W_i \sim \text{Bernoulli}(p)$ for all $i$.

Our primary objective is to estimate the ATE, defined as $\tau = \E[Y(1) - Y(0)]$. To ensure identifiability of the ATE, we assume that the potential outcomes are independent of the treatment assignment, i.e., $(Y_i(0), Y_i(1)) \indep W_i$. Under these assumptions, the ATE can be identified using the observed data: $\tau = \E[Y(1)] - \E[Y(0)] = \E[Y(1) \given W=1] - \E[Y(0) \given W=0] = \E[Y \given W=1] - \E[Y \given W=0]$.

For notation simplicity, let $n_1 = \sum_{i=1}^n W_i$ denote the number of units in the treatment group and $n_0 = \sum_{i=1}^n (1 - W_i)$ denote the number of units in the control group. The sample means of the outcomes in each group are given by $\overline Y_1 = n_1^{-1} \sum_{W_i = 1} Y_i$ and $\overline Y_0 = n_0^{-1} \sum_{W_i = 0} Y_i$. Similarly, the sample means of the pre-treatment covariates are $\overline X_1 = n_1^{-1} \sum_{W_i = 1} X_i$ and $\overline X_0 = n_0^{-1} \sum_{W_i = 0} X_i$.

\vspace{-1em}

\subsection{Review of CUPED and CUPAC}

Based on the identification $\tau = \E[Y \given W=1] - \E[Y \given W=0]$, a natural estimator for the ATE is the difference-in-means estimator, defined as $\hat\tau_{\rm DIFF} = n_1^{-1} \sum_{W_i=1} Y_i - n_0^{-1} \sum_{W_i=0} Y_i$. $\hat\tau_{\rm DIFF}$ is unbiased in finite samples, i.e., $\E[\hat\tau_{\rm DIFF}] = \tau$. The asymptotic distribution of $\hat\tau_{\rm DIFF}$ is $\sqrt{n} (\hat\tau_{\rm DIFF} - \tau) \stackrel{\sf d}{\longrightarrow} N(0,\sigma_{\rm DIFF}^2)$, where the asymptotic variance $\sigma_{\mathrm{DIFF}}^2$ is given by $\sigma_{\rm DIFF}^2 = p^{-1} \Var[Y\given W=1] + (1-p)^{-1} \Var[Y\given W=0]$. Let $\widehat\Var[Y\given W=1] = (n_1-1)^{-1} \sum_{W_i=1}(Y_i - \overline Y_1)^2$ and $\widehat\Var[Y\given W=0] = (n_0-1)^{-1}\sum_{W_i=0}(Y_i - \overline Y_0)^2$ denote the sample variances within each group. Then the variance estimator $\hat\sigma_{\rm DIFF}^2 = n(n_1^{-1} \widehat\Var[Y\given W=1] + n_0^{-1} \widehat\Var[Y\given W=0])$ is a consistent estimator of $\sigma_{\rm DIFF}^2$, i.e., $\hat\sigma_{\rm DIFF}^2$ converges in probability to $\sigma_{\rm DIFF}^2$.

To utilize pre-treatment covariates for variance reduction, the CUPED estimator exploits the fact that the treatment assignment $W$ is independent of the pre-treatment covariates $X$, i.e., $W \indep X$. This independence ensures that regression adjustments using $X$ do not introduce bias into the estimator. CUPED applies linear regression adjustment to reduce variance. CUPED is of the form $\hat\tau_{\rm CUPED} = n_1^{-1} \sum_{W_i=1} (Y_i - \hat\theta^\top X_i) - n_0^{-1} \sum_{W_i=0} (Y_i - \hat\theta^\top X_i)$, where $\hat{\theta} \in \mathbb{R}^d$ is estimated by regressing $Y$ on $X$ using the entire dataset. Specifically, CUPED predicts $Y$ using a linear function of $X$: $\theta^\top X + \theta_0$, where $\theta \in \bR^d$ and $\theta_0 \in \mathbb{R}$. The optimal $\theta$ minimizing the mean squared error (MSE) is given by $\theta = (\Var[X])^{-1} \Cov[X,Y]$. The estimation error due to using $\hat{\theta}$ instead of $\theta$ is $(\theta - \hat\theta)^\top (\overline X_1 - \overline X_0)$. Since $\hat{\theta} - \theta = o_\P(1)$ due to the properties of least squares estimation, and $\overline X_1 - \overline X_0 = O_\P(n^{-1/2})$ by $\E[X \given W=1] = \E[X \given W=0]$, the estimation error is $o_\P(n^{-1/2})$, which is asymptotically negligible. Therefore, we have $\hat\tau_{\rm CUPED} = n_1^{-1} \sum_{W_i=1} (Y_i - \theta^\top X_i) - n_0^{-1} \sum_{W_i=0} (Y_i - \theta^\top X_i) + o_\P(n^{-1/2})$. Although CUPED is not generally unbiased in finite samples without sample splitting, the consistency can be guaranteed by $\E[X \given W=1] = \E[X \given W=0]$. The asymptotic distribution of CUPED is $\sqrt{n} (\hat\tau_{\rm CUPED} - \tau) \stackrel{\sf d}{\longrightarrow} N(0,\sigma_{\rm CUPED}^2)$, where the asymptotic variance $\sigma_{\mathrm{CUPED}}^2$ is $\sigma_{\rm CUPED}^2 = p^{-1} \Var[Y - \theta^\top X \given W=1] + (1-p)^{-1} \Var[Y - \theta^\top X \given W=0]$. We estimate the variances within each group as $\widehat\Var[Y - \theta^\top X \given W=1] = (n_1-1)^{-1} \sum_{W_i=1}(Y_i - \hat\theta^\top X_i - \overline Y_1 + \hat\theta^\top \overline X_1)^2$ and $\widehat\Var[Y - \theta^\top X \given W=0] = (n_0-1)^{-1} \sum_{W_i=0}(Y_i - \hat\theta^\top X_i - \overline Y_0 + \hat\theta^\top \overline X_0)^2$. Then the variance estimator is $\hat\sigma_{\rm CUPED}^2 = n(n_1^{-1}\widehat\Var[Y - \theta^\top X\given W=1] + n_0^{-1} \widehat\Var[Y - \theta^\top X \given W=0])$, which is consistent for $\sigma_{\mathrm{CUPED}}^2$.

The CUPAC estimator generalizes CUPED by allowing for nonlinear models in the regression adjustment. Instead of restricting to linear functions, CUPAC uses a potentially complex function $f(X)$ to predict $Y$. The estimator is defined as $\hat\tau_{\rm CUPAC} = n_1^{-1} \sum_{W_i=1} (Y_i - \hat f(X_i)) - n_0^{-1} \sum_{W_i=0} (Y_i - \hat f(X_i))$, where $\hat{f}$ is an estimate of the optimal function $f$ obtained by fitting a machine learning model to predict $Y$ from $X$. We assume $\hat f$ is $L_2$-consistent in estimating $f$. The estimation error in this case is $n_1^{-1} \sum_{W_i=1} (f(X_i) - \hat f(X_i)) - n_0^{-1} \sum_{W_i=0} (f(X_i) - \hat f(X_i))$. By the minimax theory, the pointwise convergence rate of $\hat f$ estimating $f$ is generally slower than $n^{-1/2}$. However, since $X \given W=1$ and $X \given W=0$ have the same distribution, under the Donsker conditions, as long as $\hat f$ is $L_2$ consistent, the estimation error is $o_\P(n^{-1/2})$. This assumption allows us to avoid using sample splitting. Therefore, we have $\hat\tau_{\rm CUPAC} = n_1^{-1} \sum_{W_i=1} (Y_i - f(X_i)) - n_0^{-1} \sum_{W_i=0} (Y_i - f(X_i)) + o_\P(n^{-1/2})$. Similar to CUPED, CUPAC is not unbiased in finite samples without sample splitting, but it remains consistent because $W \indep X$ implies that $\E[f(X) \given W=1] = \E[f(X) \given W=0]$ for any measurable function $f$. The asymptotic distribution of CUPAC is $\sqrt{n} (\hat\tau_{\rm CUPAC} - \tau) \stackrel{\sf d}{\longrightarrow} N(0,\sigma_{\rm CUPAC}^2)$, where the asymptotic variance $\sigma_{\mathrm{CUPAC}}^2$ is $\sigma_{\rm CUPAC}^2 = p^{-1} \Var[Y - f(X) \given W=1] + (1-p)^{-1} \Var[Y - f(X) \given W=0]$. We estimate the variances within each group as $\widehat\Var[Y - f(X) \given W=1] = (n_1-1)^{-1} \sum_{W_i=1}(Y_i - \hat f(X_i) - \overline Y_1 + n_1^{-1} \sum_{W_i=1} \hat f(X_i))^2$ and $\widehat\Var[Y - f(X) \given W=0] = (n_0-1)^{-1} \sum_{W_i=0}(Y_i - \hat f(X_i) - \overline Y_0 + n_0^{-1} \sum_{W_i=0} \hat f(X_i))^2$. Then the variance estimator $\hat\sigma_{\rm CUPAC}^2 = n(n_1^{-1} \widehat\Var[Y - f(X)\given W=1] + n_0^{-1} \widehat\Var[Y - f(X) \given W=0])$ is consistent for $\sigma_{\mathrm{CUPAC}}^2$.

\subsection{Related work}

This paper considers the ATE estimation problem in randomized experiment setting, a topic with a rich history in causal inference \citep{imbens2015causal, ding2024first}. To improve the efficiency of the estimator using regression adjustment with pre-experiment data, \cite{freedman2008regression} and \cite{lin2013agnostic} explored linear models. \cite{cohen2024no}, \cite{guo2021machine} and \cite{jin2023toward} investigated nonlinear models with sample splitting and cross fitting, which is closely related to double machine learning \citep{chernozhukov2018double}. Specifically, \cite{jin2023toward} focused on optimal variance reduction by using separate outcome models for the treatment and control groups. In our proposed method, similar to CUPED and CUPAC, we utilize a single model for both groups, although all subsequent results are applicable to the two-model case as well. The CUPAC estimator we consider, as introduced by \cite{tang2020control}, is implemented without sample splitting, akin to the methods used by \cite{lin2023estimation, lin2025regression, cattaneo2025rosenbaum} for estimating the ATE using observational data. When flexible nuisance models are trained on the same data used for treatment effect estimation, sample splitting and cross-fitting is a standard way to avoid restrictive empirical process conditions. We focus, however, on the practically common industry pipeline in which the prediction model is trained offline on historical data and then deployed to current experiments. In that setting the first-stage predictor is effectively fixed with respect to the experimental sample, and the empirical process conditions become unnecessary. We state results without sample splitting under empirical process conditions for completeness.

Recent studies have increasingly focused on using in-experiment data, or post-treatment covariates, to improve the estimation of the ATE. One popular method involves using short-term surrogates to estimate unobservable long-term treatment effects; see \cite{athey2025surrogate} and references therein. \cite{deng2023variance} applied surrogate methods specifically to estimate delayed outcomes in online experimental settings. A key assumption in surrogate methods is that the treatment and the outcome of interest are conditionally independent given the surrogates \citep{prentice1989surrogate}, which is strong and not testable in practice \citep{vanderweele2013surrogate}. Additionally, some works have focused on principal stratification approaches, which assume principal ignorability \citep{frangakis2002principal, ding2017principal, jiang2022multiply}, which are also untestable due to the fundamental problem of causal inference. \cite{deng2023zero} incorporated the idea of principal stratification and instrumental variables into the CUPED framework. In contrast to these approaches, we propose an alternative way that selects and utilizes post-treatment covariates that are plausibly treatment-insensitive for the intervention under study. The assumptions in our method are testable, as we focus on different sets of post-treatment covariates compared to those used in surrogate and principal stratification methods. Specifically, our framework targets post-treatment covariates that behave analogously to pre-treatment covariates, rather than classical mediators or surrogates. Such post-treatment covariates are frequently available in practical A/B testing scenarios but are often disregarded due to concerns about introducing post-treatment bias.

\section{Method}\label{sec:method}
 
\subsection{Motivation}

In this section, we compare the asymptotic variances of the three methods, difference-in-means, CUPED, and CUPAC—to understand the extent of variance reduction achieved by CUPED and CUPAC, thereby motivating our proposed approach. The asymptotic variances of these estimators depend linearly on the variances of their residuals: $\Var[Y \given W]$ for the difference-in-means, $\Var[Y - \theta^\top X \given W]$ for CUPED, and $\Var[Y - f(X) \given W]$ for CUPAC. This relationship indicates that the amount of uncertainty in $Y$ explained by $X$ directly influences the proportion of variance reduction attainable by CUPED and CUPAC. When $Y$ and $X$ are strongly correlated, the residual variance is low, leading to substantial variance reduction. Conversely, when $Y$ and $X$ are independent, we have $\theta = 0$ for CUPED and $f(X) = \E[Y]$ for CUPAC, resulting in no variance reduction. This observation motivates the incorporation of in-experiment data, as such data are generally more closely related to the outcome and can potentially lead to greater variance reduction.

To understand the sources of uncertainty in $Y$, consider an unobserved variable $U$ representing inherent randomness or unmeasured factors affecting $Y$. Fixing the treatment assignment $W$, we examine the relationships among $U$, $X$, $Y$, and $W$, as depicted in Figure~\ref{fig1}.

\begin{figure}[h]
    \centering
    \subfigure[Relationships among $U$, $X$, $Y$, $W$.]{
        \begin{tikzpicture}[->,>=stealth',shorten >=1pt,auto,node distance=2cm, thick, scale=1.1, transform shape]
            \node (Y) {$Y$};
            \node (W) [right of=Y] {$W$};
            \node (X) [left of=Y] {$X$};
            \node (U) [below of=Y] {$U$};
            \path (W) edge[ultra thick] (Y)
                  (X) edge[dashed] (Y)
                  (U) edge (Y)
                  (U) edge (X);
        \end{tikzpicture}
        \label{fig1}
    }
    \hspace{2cm}
    \subfigure[Relationships among $U$, $X$, $Y$, $W$, $Z$.]{
        \begin{tikzpicture}[->,>=stealth',shorten >=1pt,auto,node distance=2cm, thick, scale=1.1, transform shape]
            \node (Y) {$Y$};
            \node (W) [right of=Y] {$W$};
            \node (X) [left of=Y] {$X$};
            \node (U) [below of=Y] {$U$};
            \node (Z) [below of=W] {$Z$};
            \path (W) edge[ultra thick] (Y)
                  (W) edge[ultra thick] (Z)
                  (Z) edge[dashed] (Y)
                  (X) edge[dashed] (Y)
                  (U) edge (Y)
                  (U) edge (X)
                  (U) edge (Z);
        \end{tikzpicture}
        \label{fig2}
    }
    \caption{Comparison of variable relationships in two causal models.}
\end{figure}

In Figure~\ref{fig1}, the bold arrow from $W$ to $Y$ represents the treatment effect of interest. The uncertainty in $Y$, given fixed $W$, arises from two sources. The first is the direct effect of $U$ on $Y$, representing random errors or unobserved covariates not captured by $X$. The second is the indirect effect of $U$ on $Y$ through $X$, since $Y$ depends on $X$ via the outcome model, and $X$ itself varies due to random sampling from its marginal distribution. By regressing $Y$ on $X$, we effectively remove the dashed arrow from $X$ to $Y$, thereby blocking the indirect path from $U$ to $Y$ through $X$. This reduces the variance of the estimator by eliminating the variability in $Y$ that can be explained by $X$.

Now, we consider incorporating in-experiment data into the method. Suppose we have i.i.d. samples $\{(W_i, X_i, Z_i, Y_i(0), Y_i(1))\}_{i=1}^n$ drawn from the joint distribution $(W, X, Z, Y(0), Y(1))$. Here, $Z \in \mathbb{R}^m$ represents post-treatment covariates, which are observed after the treatment assignment and may depend on $W$. This means the treatment may influence the outcome through $Z$. The relationships among $U$, $X$, $Y$, $W$, and $Z$ are illustrated in Figure~\ref{fig2}.

In Figure~\ref{fig2}, the treatment effect consists of two components: the direct effect from $W$ to $Y$, and the indirect effect from $W$ to $Y$ mediated through $Z$. The uncertainty in $Z$ arises from random sampling of its conditional distribution given $W$. If we treat $Z$ similarly to $X$ and apply regression adjustment using $Z$, we remove the dashed arrow from $Z$ to $Y$, which blocks the indirect path from $W$ to $Y$ through $Z$. This would eliminate part of the treatment effect, resulting in a biased estimator of the ATE.

Figure~\ref{fig2} depicts a generic post-treatment variable. Our method does not adjust for such variables indiscriminately. To ensure that we can reduce the uncertainty from $U$ to $Y$ through $Z$ without introducing bias, we need to carefully evaluate the role of $Z$ in the causal pathway. Specifically, we must determine whether the indirect effect from $W$ to $Y$ through $Z$ exists and whether adjusting for $Z$ is appropriate in this context. By incorporating a subset of post-treatment covariates that are plausibly treatment-insensitive for the intervention under study, we can leverage their stronger correlation with the outcome to achieve greater variance reduction in the estimation of the ATE.

\subsection{Method overview}

In the context of online controlled experiments, the treatment may influence the outcome only through a subset of post-treatment covariates. Our key idea for combining pre-experiment and in-experiment data is to identify post-treatment covariates that have no indirect effect on the outcome and incorporate them into the CUPAC framework via a second-stage linear adjustment. Importantly, the assumption of no indirect effect does not require that these covariates are independent of the treatment. Instead, we rely on weaker and more practical assumptions. By the linear adjustment, it suffices to assume mean equivalence of the selected post-treatment covariates across treatment arms. This assumption is less restrictive than requiring full independence and is empirically testable. In contrast, the nonlinear adjustment assumes distributional equivalence, which is stronger and requires independence testings \cite{lin2023boosting, lin2022limit,lin2024failure}. Consequently, our method uses a second-stage linear adjustment, leveraging its weaker assumption to enable the inclusion of a broader set of post-treatment covariates while reducing the risk of incorporating covariates that introduce post-treatment bias.

Let $Z(w)$ denote the value of a candidate post-treatment covariate $Z$ under treatment arm $w \in \{0,1\}$. A classical mediator is a variable for which $Z(1)\neq Z(0)$ and whose change carries part of the treatment effect to the outcome. Such variables should not be adjusted for. The strongest sufficient condition for safe adjustment is $Z(1)=Z(0)$ almost surely, meaning that the intervention does not affect the covariate at all. For the linear second-stage adjustment considered here, however, unbiasedness only requires the weaker moment condition $\E[Z(1)] = \E[Z(0)]$, which under randomization is equivalent to $\E[Z \given W=1] = \E[Z \given W=0]$. Thus mean equivalence does not claim that the entire distribution of $Z$ is invariant; higher-order moments or nonlinear features of $Z$ may still differ across arms. Such differences do not bias a linear adjustment unless they induce a difference in the mean of the selected covariates.

We now assume that, after post-treatment covariates selection, all selected covariates $Z$ satisfy the mean equivalence condition $\E[Z \given W=1] = \E[Z \given W=0]$. To incorporate the pre-experiment data, we begin by predicting $Y$ using $X$ and obtain a fitted model $\hat f(\cdot)$ in the same manner as in CUPAC. Let the fitted residuals be $\hat R_i = Y_i - \hat f(X_i)$, which capture the variation in $Y$ that is not explained by the pre-experiment covariates. We then regress $\hat R_i$ on $Z_i$ using a linear model and obtain the coefficient vector $\hat\gamma$. Ideally, $\hat{\gamma}$ estimates $\gamma = (\Var[Z])^{-1} \Cov[Z, Y - f(X)]$. Our final estimator is defined as
\[
 \hat\tau = n_1^{-1} \sum_{W_i=1} (Y_i - \hat f(X_i) - \hat\gamma^\top Z_i) - n_0^{-1} \sum_{W_i=0} (Y_i - \hat f(X_i) - \hat\gamma^\top Z_i).
\]
The estimand of our estimator is
\[
\E[Y - f(X) - \gamma^\top Z \given W=1] - \E[Y - f(X) - \gamma^\top Z \given W=0]
= \tau - \gamma^\top\{\E[Z \given W=1]-\E[Z \given W=0]\}.
\]
Hence the bias from using $Z$ is completely determined by the mean imbalance of the selected linear covariates. If mean equivalence holds, the adjustment is unbiased. If it fails by $\Delta = \E[Z \given W=1]-\E[Z \given W=0]$, then the induced bias is $-\gamma^\top \Delta$. This formula clarifies both the strength and the limit of our safeguard: mild violations matter only to the extent that the imbalanced covariates receive substantial regression weight.

The estimation error from the first step is the same as in CUPAC, $n_1^{-1} \sum_{W_i=1} (f(X_i) - \hat f(X_i)) - n_0^{-1} \sum_{W_i=0} (f(X_i) - \hat f(X_i))$, which is $o_\P(n^{-1/2})$ under certain conditions. Let $\overline Z_1 = n_1^{-1} \sum_{W_i = 1} Z_i$ and $\overline Z_0 = n_0^{-1} \sum_{W_i = 0} Z_i$ be the sample means of the selected post-treatment covariates in each group. The estimation error from the second step is $(\hat\gamma - \gamma)^\top (\overline Z_1 - \overline Z_0)$. Since $\E[Z \given W=1] = \E[Z \given W=0]$, we have $\overline Z_1 - \overline Z_0 = O_\P(n^{-1/2})$. Although we do not have access to the oracle residuals $Y_i - f(X_i)$, we can still establish the following proposition, provided that $\hat{f}$ estimates $f$ accurately.
\begin{proposition}\label{prop:1}
    We have $\hat\gamma - \gamma = O_\P(\lVert \hat f - f \rVert_2) + o_\P(1)$.
\end{proposition}

As long as $\hat{f}$ is $L_2$-consistent in estimating $f$, the second estimation error is $o_\P(n^{-1/2})$. Therefore, our estimator becomes
\[
\hat\tau = \frac{1}{n_1} \sum_{W_i=1} (Y_i - f(X_i) - \gamma^\top Z_i) - \frac{1}{n_0} \sum_{W_i=0} (Y_i - f(X_i) - \gamma^\top Z_i) + o_\P(n^{-1/2}).
\]
The estimator is consistent with asymptotic distribution $\sqrt{n} (\hat\tau - \tau) \stackrel{\sf d}{\longrightarrow} N(0,\sigma^2)$, where
\begin{align*}
    \sigma^2 = p^{-1} \Var[Y - f(X) - \gamma^\top Z \given W=1] +  (1-p)^{-1} \Var[Y - f(X) - \gamma^\top Z \given W=0].
\end{align*}

Because treatment is randomized and the propensity score is known, these validity results do not require correct specification of either the first-stage prediction model or the second-stage linear projection. A full semiparametric efficiency analysis is beyond the scope of this paper. However, in the benchmark case where the combined adjustment $f(X)+\gamma^\top Z$ coincides with the true conditional mean outcome, the resulting adjusted difference-in-means estimator attains the semiparametric efficiency bound for randomized experiments. Without that additional condition, the estimator remains valid and can still reduce variance substantially, but it need not be efficient.

Let us define the estimated variances within each group $\widehat\Var[Y - f(X) - \gamma^\top Z \given W=1] = (n_1-1)^{-1} \sum_{W_i=1}(Y_i - \hat f(X_i) - \hat\gamma^\top Z_i - \overline Y_1 + n_1^{-1} \sum_{W_i=1} \hat f(X_i) + \hat\gamma^\top \overline Z_1)^2$ and $\widehat\Var[Y - f(X) - \gamma^\top Z \given W=0] = (n_0-1)^{-1} \sum_{W_i=0}(Y_i - \hat f(X_i) - \hat\gamma^\top Z_i - \overline Y_0 + n_0^{-1} \sum_{W_i=0} \hat f(X_i) + \hat\gamma^\top \overline Z_0)^2$. Then the consistent variance estimator for $\sigma^2$ is
\begin{align*}
 \hat\sigma^2 = (n/n_1) \widehat\Var[Y - f(X) - \gamma^\top Z\given W=1] + (n/n_0) \widehat\Var[Y - f(X) - \gamma^\top Z \given W=0].
\end{align*}

\subsection{An example}

To illustrate the level of variance reduction achieved by our method, consider an additive model for the outcome variable $Y$: $Y = g(X) + h(Z) + \tau W + \epsilon$. Here, $g(\cdot): \bR^d \to \bR$ and $h(\cdot): \bR^m \to \bR$ are functions of the pre-treatment covariates $X$ and the post-treatment covariates $Z$, respectively. We assume that $X$ and $Z$ are standardized such that $\Var[X] = I_d$ and $\Var[Z] = I_m$, where $I_d$ and $I_m$ are identity matrices of dimensions $d$ and $m$. Let the treatment assignment probability be $p = \P(W=1) = 1/2$. We also assume that $\E[\epsilon] = 0$ and $\Var[\epsilon] = \sigma_{\epsilon}^2$, representing the variance of the exogenous noise $\epsilon$, which is independent of $X$ and $Z$. Furthermore, due to the randomization in the experiment, $(X,Z,\epsilon)$ is independent of $W$.

To predict $Y$ using only the pre-treatment covariates $X$, the optimal model under the MSE risk is $\E[Y \given X] = g(X) + \E[h(Z) \given X] + \tau p$. We assume our function class for the prediction model is sufficiently rich to include $\E[Y \given X]$. Therefore, we set $f(X) = g(X) + \E[h(Z) \given X] + \tau p$. The coefficient $\gamma$ used in our method is given by $\gamma = (\Var[Z])^{-1} \Cov[Z, Y - f(X)] = \Cov[Z, h(Z) + \tau W + \epsilon - \E[h(Z) \given X] - \tau p] = \Cov[Z, h(Z) - \E[h(Z) \given X]] = \E[\Cov[Z, h(Z)\given X]]$.

Next, we compute the asymptotic variances of the different estimators. The asymptotic variance of the difference-in-means estimator is $\sigma_{\mathrm{DIFF}}^2 = \Var[g(X)+h(Z)+\epsilon] = \Var[g(X)+h(Z)] + \sigma_{\epsilon}^2 = \Var[g(X) + \E[h(Z)\given X]] + \E[\Var[h(Z) \given X]] + \sigma_{\epsilon}^2$. For the CUPAC estimator, the asymptotic variance is $\sigma_{\rm CUPAC}^2 = \Var[g(X) + h(Z) + \epsilon - f(X)] = \Var[h(Z) - \E[h(Z) \given X] + \epsilon] = \Var[h(Z) - \E[h(Z) \given X]] + \sigma_{\epsilon}^2 = \E[\Var[h(Z) \given X]] + \sigma_{\epsilon}^2$. For our proposed estimator, the asymptotic variance is $\sigma^2 = \Var[g(X) + h(Z) + \epsilon - f(X) - \gamma^\top Z] = \Var[h(Z) - \E[h(Z) \given X] - \gamma^\top Z + \epsilon] = \Var[h(Z) - \E[h(Z) \given X] - \gamma^\top Z] + \sigma_{\epsilon}^2 = \E[\Var[h(Z) \given X]] - \lVert \E[\Cov[Z, h(Z)\given X]] \rVert_2^2 + \sigma_{\epsilon}^2$.

From the above calculations, the variance reduction achieved by CUPAC over the difference-in-means estimator is $\sigma_{\mathrm{DIFF}}^2 - \sigma_{\rm CUPAC}^2 = \Var[g(X) + \E[h(Z)\given X]]$, which represents the variance in $Y$ that can be explained by the pre-treatment covariates $X$. The additional variance reduction achieved by our method over CUPAC is $\sigma_{\rm CUPAC}^2 - \sigma^2 = \lVert \E[\Cov[Z, h(Z)\given X]] \rVert_2^2$, capturing the variance not explained by $X$ but explainable by a linear combination of $Z$.

% An important observation is that both $\sigma_{\mathrm{DIFF}}^2 - \sigma_{\rm CUPAC}^2$ and $\sigma_{\rm CUPAC}^2 - \sigma^2$ are independent of the noise variance $\sigma_{\epsilon}^2$ and depend only on the relationships between $X$, $Z$, and the functions $g$ and $h$. This is intuitive since regression adjustments cannot reduce variance due to exogenous noise $\epsilon$. Consequently, the proportion of variance reduction $1-\sigma_{\rm CUPAC}^2/\sigma_{\mathrm{DIFF}}^2$ and $1-\sigma^2/\sigma_{\rm CUPAC}^2$ depends on the relative magnitude of the noise variance. If the noise variance $\sigma_\epsilon^2$ is large, the proportion of variance reduction achievable through regression adjustment decreases.

To understand the effect of the correlation between $X$ and $Z$ on variance reduction, consider two scenarios. First, if $X$ and $Z$ are independent, the variance reduction by CUPAC is $\sigma_{\mathrm{DIFF}}^2 - \sigma_{\rm CUPAC}^2 = \Var[g(X)]$  which is the variance explained by $X$ alone. The additional variance reduction achieved by our method over CUPAC is: $\sigma_{\rm CUPAC}^2 - \sigma^2 = \lVert \Cov[Z, h(Z)] \rVert_2^2$. Here $\sigma_{\mathrm{DIFF}}^2 - \sigma_{\rm CUPAC}^2 = \Var[g(X)]$ is a quadratic functional of $g$, and $\sigma_{\rm CUPAC}^2 - \sigma^2 = \lVert \Cov[Z, h(Z)] \rVert_2^2$ is a quadratic functional of $h$. Since $Z$ is more strongly correlated with the outcome than $X$, we can imagine that $h$ has a larger scale than $g$. Consequently, $\sigma_{\rm CUPAC}^2 - \sigma^2$ will be larger than $\sigma_{\mathrm{DIFF}}^2 - \sigma_{\rm CUPAC}^2$. For example, consider linear functions $g(X) = \beta_g^\top X$ and $h(Z) = \beta_h^\top Z$, where $\beta_g \in \bR^d$ and $\beta_h \in \bR^m$. The vectors $\beta_g$ and $\beta_h$ represent the sizes of the effects of $X$ and $Z$ on the outcome, respectively. Then we have $\sigma_{\mathrm{DIFF}}^2 - \sigma_{\rm CUPAC}^2 = \lVert \beta_g \rVert_2^2$ and $\sigma_{\rm CUPAC}^2 - \sigma^2 = \lVert \beta_h \rVert_2^2$. This indicates that we can achieve a larger variance reduction with our method if the effect of the post-treatment covariates $Z$ is large. On the other hand, if $X$ and $Z$ are strongly correlated such that $\E[h(Z) \given X] = h(Z)$ for almost all $X$, then $\sigma_{\mathrm{DIFF}}^2 - \sigma_{\rm CUPAC}^2 = \Var[g(X) + h(Z)]$ and $\sigma_{\rm CUPAC}^2 - \sigma^2 = 0$. In this case, our method does not provide additional variance reduction since $X$ and $Z$ share the same information, leaving no residual variance after using CUPAC.

Next, we consider the effect of the complexity of the underlying model $h$ of $Z$ on variance reduction. The variance reduction achieved by CUPAC over the difference-in-means estimator, $\sigma_{\mathrm{DIFF}}^2 - \sigma_{\rm CUPAC}^2$, depends on $\E[h(Z)\given X]$. Therefore, the complexity of the model $h$ has no effect on this variance reduction as long as the conditional mean remains the same. However, the additional variance reduction achieved by our method over CUPAC, $\sigma_{\rm CUPAC}^2 - \sigma^2$, depends on $\Cov[Z, h(Z)\given X]$. This indicates that the level of variance reduction depends on the projection of $h(Z) \given X$ onto the space spanned by $Z \given X$. For instance, if $h(Z) = \beta_h^\top Z$ for some $\beta_h \in \bR^m$, we achieve the maximum variance reduction because the relationship between $h(Z)$ and $Z$ is linear and directly captured by $Z$. Conversely, if $h(Z)\given X$ is orthogonal to the space spanned by $Z \given X$ for almost all $X$, there is no variance reduction from our method.

\subsection{Post-treatment covariates selection}

In practice, not all post-treatment covariates will satisfy the assumption required by our method, namely mean equivalence across treatment and control groups. This is because the treatment may influence the outcome through a subset of these covariates. Therefore, it is crucial to identify which candidates can be safely used in the second-stage adjustment without introducing bias.

To identify suitable covariates from a set of candidates $Z = (Z^{(1)}, Z^{(2)}, \ldots, Z^{(m)})$, we employ two-sample statistical tests for each individual post-treatment covariate. Specifically, for each covariate $Z^{(j)}$, where $j \in \{1, 2, \ldots, m\}$, we test the null hypothesis $H_0: \E[Z^{(j)} \given W=1] = \E[Z^{(j)} \given W=0]$ using data from both treatment and control groups. Let $Z_i = (Z_{i1}, Z_{i2}, \ldots, Z_{im})$ denote the observed values of the covariates for unit $i$, where $i = 1, \ldots, n$. For each covariate $Z^{(j)}$, we perform a two-sample test using the observations $\{Z_{ij}: W_i=1\}$ from the treatment group and $\{Z_{ij}: W_i=0\}$ from the control group, and obtain the corresponding p-value $p_j$. Given a significance level $\alpha \in (0,1)$, we select the covariates for which the null hypothesis cannot be rejected, i.e., those with $p_j > \alpha$. Importantly, this non-rejection should be interpreted as a diagnostic rather than a proof that the null is true: a large p-value may reflect either genuine balance or limited power. The screening step is therefore not intended to be the sole safeguard and should be combined with domain knowledge about whether a selected covariate is plausibly treatment-insensitive.

We show that, asymptotically, all selected post-treatment covariates satisfy the mean equivalence assumption, or equivalently, any covariates that violate the assumption will almost surely not be selected.
\begin{proposition}\label{prop:2}
    Let $\mathcal{S}_0 \subset \{1, 2, \ldots, m\}$ denote the indices of covariates satisfying $H_0$, and let $\mathcal{S}_1 \subset \{1, 2, \ldots, m\}$ denote the indices of covariates violating $H_0$. Consider a significance level $\alpha \in (0,1)$. Assume that the two-sample tests used are consistent. Then,
    \[
        \lim_{n \to \infty}\P(p_j \le \alpha, \forall j \in \cS_1) = 1.
    \]
\end{proposition}
 
Proposition~\ref{prop:2} holds for any choice of $\alpha \in (0,1)$, provided that the tests are consistent. While Proposition~\ref{prop:2} ensures that the selected covariates satisfy the mean equivalence assumption asymptotically, a natural question arises: Can we guarantee that all covariates satisfying the mean equivalence assumption are included in our selection? This issue relates to controlling the family-wise error rate (FWER), which is the probability of making one or more Type I errors when performing multiple hypothesis tests, i.e., $\P(p_j \le \alpha, \exists j \in \cS_0)$. To address this, we can apply multiple testing correction procedures, such as the Bonferroni correction or the Holm-Bonferroni method, to adjust the p-values and control the FWER at a desired level. Furthermore, if we desire the FWER to approach zero, we can let $\alpha$ tend to zero, allowing us to correctly identify $\cS_0$ in Proposition~\ref{prop:2} with probability one asymptotically, thereby achieving selection consistency. We show this fact in Proposition~\ref{prop:3}. See \cite{shi2025forward} and the references therein for further discussions on selection consistency.
\begin{proposition}\label{prop:3}
    Using the same notation as in Proposition~\ref{prop:2}, and assuming that the two-sample tests used are consistent for any $\alpha \in (0,1)$, we have
    \[
        \lim_{\alpha \to 0} \lim_{n \to \infty}\P(\{j:p_j > \alpha\} = \cS_0) = 1.
    \]
\end{proposition}

Both Proposition~\ref{prop:2} and Proposition~\ref{prop:3} address the asymptotic behavior of the selection procedure, which is particularly relevant given that online experiments are typically conducted at large scale with large sample sizes. In finite samples, however, the behavior of the screening step depends on power, multiplicity, and the practical tolerance for imbalance. The finite-sample behavior depends on the chosen significance level $\alpha$ and the magnitude of the differences in means between treatment and control groups for each covariate. By employing efficient two-sample tests, even small deviations from the null can be detected under moderate sample sizes. Empirically, online experiments generally involve large sample sizes, so standard two-sample tests often exhibit high power to detect violations of mean equivalence. At the same time, when the sample size is very large, even practically negligible mean differences may become statistically significant. This is one reason to complement hypothesis testing with effect-size diagnostics and substantive screening.

In addition to traditional hypothesis testing for mean equivalence, an alternative approach is to use equivalence testing \citep{westlake1972use, hauck1984new} to select post-treatment covariates. After specifying equivalence bounds, one can implement the Two One-Sided Tests procedure \citep{schuirmann1987comparison, walker2011understanding}, selecting the covariate only if both one-sided tests are significant. See \cite{lakens2017equivalence} for practical guidance on applying equivalence testing. This approach can be particularly attractive when sample sizes are so large that tiny but practically irrelevant differences are routinely detectable.

When the number of candidate covariates is large, strong family-wise error-rate control via Bonferroni or Holm adjustment can become conservative. A practical workflow is to first use domain knowledge to exclude obvious mediators, then aggregate evidence across historical experiments when possible, and finally apply either strong multiplicity correction for confirmatory deployment or equivalence tests with practically meaningful bounds. The trade-off is asymmetric: false inclusion can introduce bias, whereas false exclusion only sacrifices efficiency. For this reason, conservative selection is often preferable in production experimentation systems. Following selection, practitioners can further validate the robustness of chosen covariates by inspecting empirical distributions and by checking whether the selected set remains stable across related experiments.
 
\subsection{Practical issues}

One practical advantage of using in-experiment data over pre-experiment data is its universal availability—unlike pre-experiment data, which may be missing for a subset of users. For example, in A/B tests involving new users, no historical data exist to support variance reduction. Moreover, access to pre-experiment data may be restricted by evolving privacy policies or constrained by limited data retention windows. In contrast, in-experiment data are collected during the experiment itself and are therefore available for all participants, eliminating the need to address missing data issues.

In many companies, the sample sizes for A/B testing are very large, and multiple experiments may be running simultaneously on the platform. Training a separate prediction model for each experiment, as required in CUPAC, can be both time-consuming and resource-intensive. A common practice is to assume homogeneity across different experiments regarding the prediction model. Thus, a single prediction model is pre-trained using a large dataset that combines data from past experiments, utilizing the same set of pre-treatment covariates, and then applied to all current experiments.

If the same experimental sample is used to fit $\hat f$, cross-fitting can be used in place of the Donsker-type argument above and can be combined with the same second-stage linear adjustment. In the large-scale experimentation pipelines that motivate this paper, however, the first-stage model is often trained offline on historical data and updated periodically. Once such a pretrained model is deployed, it is independent of the current experimental sample and behaves as a fixed prediction function for inference. In that common setting, the first-stage remainder is handled without requiring a Donsker condition.

To align with this production pipeline, we aim to preselect the same set of post-treatment covariates for all experiments using data from past experiments. Since the distribution of post-treatment covariates may differ across experiments, we conduct the statistical tests separately for each experiment. Because covariates may have different p-values in different experiments, we can combine the p-values across experiments using meta-analysis, e.g., Fisher's method. This approach improves the accuracy of the covariates selection procedure and allows us to identify a stable set of treatment-insensitive candidates using data from multiple past experiments.

As a generalization of our method, one might consider incorporating in-experiment data using nonlinear models instead of the linear models we use, or constructing a single prediction model using both pre-experiment and in-experiment data together as inputs. Both generalizations are valid. However, the two-step procedure has two important advantages. First, the linear second stage only requires mean equivalence of the selected covariates, whereas a flexible nonlinear second stage would generally require a stronger invariance condition. Second, by keeping the first stage identical to CUPAC, we can isolate the incremental gains from incorporating in-experiment covariates and preserve compatibility with existing experimentation pipelines.
 
\section{Empirical studies}\label{sec:exp}

In this section, we apply our proposed method to 29 online experiments conducted at Etsy over the course of one month, each involving a binary treatment. The primary outcome of interest is the customer conversion rate. For the CUPAC pipeline, we utilized 117 pre-treatment covariates and built a single prediction model using LightGBM, trained on combined data from all experiments. We use CUPAC as the primary baseline because the goal is to isolate the incremental value of adding in-experiment covariates on top of the standard production regression-adjustment pipeline.

For the selection of post-treatment covariates, the candidate variables were all count data with a large proportion of zeros across experiments. We first removed variables that were too sparse to be informative in a stable way across all experiments. We then employed the Mann–Whitney U test within each experiment and combined the resulting p-values across experiments using Fisher's method. The purpose of this screening step was to rule out clearly treatment-affected variables and retain a stable platform-level set of approximately balanced candidates rather than to prove treatment-insensitivity from p-values alone. This procedure yielded 23 post-treatment covariates for the second-stage adjustment. We did not additionally apply a formal FWER correction in this empirical study; because a covariate had to remain approximately balanced after meta-analysis across 29 experiments, the resulting screen was already conservative in practice and retained only a small subset of candidates. In larger candidate libraries or confirmatory deployments, Holm or Bonferroni adjustment can be added.

We evaluate the effectiveness of our method using two metrics that capture the level of variance reduction. The first metric is the improvement in predictive accuracy, measured by the difference in the square roots of the R-squared values between our method and the CUPAC estimator. This quantity, inspired by the asymptotic variance of the ATE estimator, reflects the gain in predictive accuracy. The second metric is the ratio of asymptotic variances, estimated using the consistent variance estimators introduced earlier.

\begin{figure}[h]
  \centering
  \includegraphics[width=\linewidth]{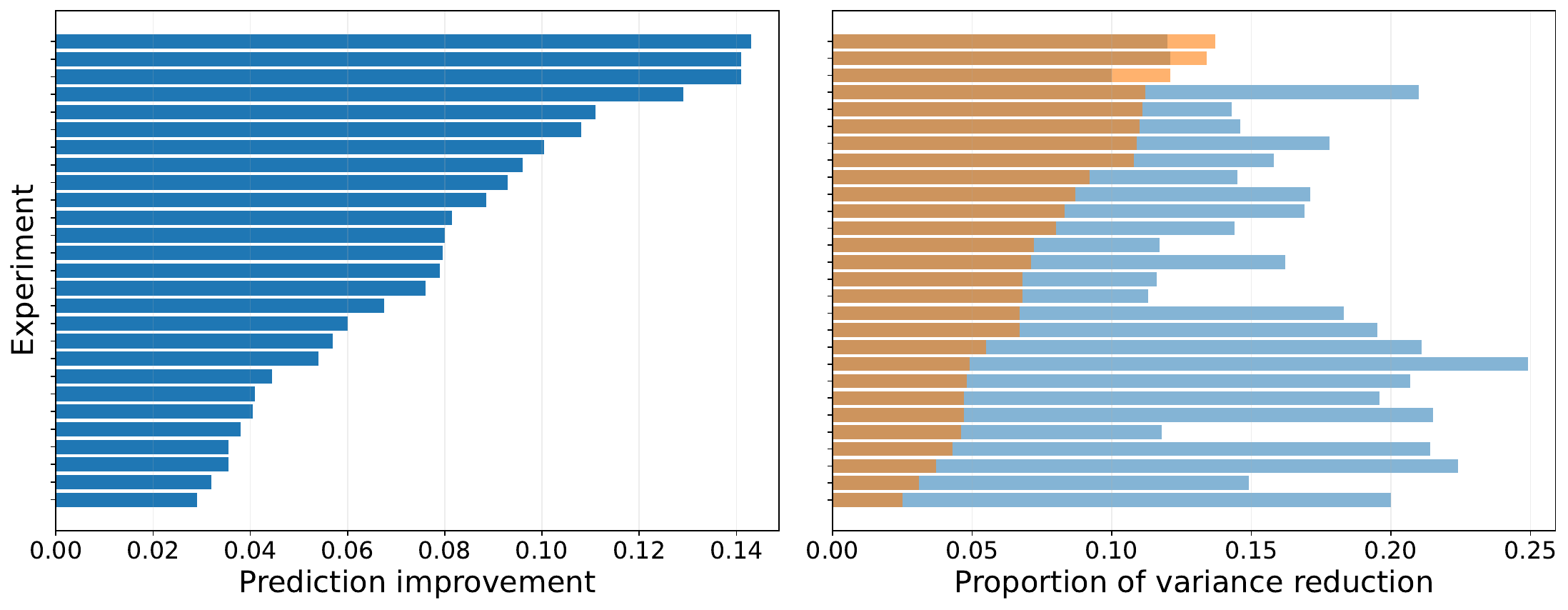}
  \caption{Variance reduction across experiments}
  \label{fig:combined}
\end{figure}

Figure~\ref{fig:combined} summarizes the variance reduction achieved across multiple experiments. Each bar on the y-axis represents a separate experiment (anonymized for privacy), while the x-axis displays the corresponding metric. Left panel shows the first metric, calculated as $\sqrt{R^2} - \sqrt{R^2_{\text{CUPAC}}}$, where $R^2$ is the R-squared from the combined model $\hat{f}(X) + \hat{\gamma}^\top Z$ and $R^2_{\text{CUPAC}}$ is the R-squared from CUPAC’s predictor $\hat{f}(X)$ alone. The results show that the improvement ranges from 0.02 to over 0.14, highlighting consistent gains in predictive performance. Right panel presents the second metric: the blue bars represent the variance reduction of CUPAC over the difference-in-means estimator, computed as $1 - \sigma_{\text{CUPAC}}^2 / \sigma_{\text{DIFF}}^2$, while the orange bars show the additional reduction achieved by our method over CUPAC, computed as $1 - \sigma^2 / \sigma_{\text{CUPAC}}^2$. These results demonstrate that our method consistently achieves comparable or greater variance reduction, despite using substantially fewer covariates—only 23 post-treatment covariates compared to the 117 pre-treatment covariates employed in CUPAC.

\section{Discussions}\label{sec:disc}

This paper introduces a general, practitioner-ready framework that leverages both pre-experiment and in-experiment data to reduce the variance of treatment effect estimates in online experiments. A key novelty of our approach is the recognition that many in-experiment covariates, though measured post-treatment, are either substantively unaffected by treatment or empirically balanced across treatment arms. Moreover, these covariates are often far more predictive of contemporaneous outcomes than pre-experiment variables, and unlike historical features, they are universally available even for users who are new to the platform. Despite their potential, industry practice typically excludes in-experiment covariates from regression adjustment due to concerns about post-treatment bias. Our results show that this exclusion is unnecessarily restrictive.

We provide a systematic, theoretically justified procedure for identifying which in-experiment covariates satisfy the required mean equivalence (or no effect) conditions. This selection step combines domain knowledge with formal balance diagnostics, enabling practitioners to safely incorporate informative in-experiment covariates without introducing bias. The practical message is not to “use all in-experiment covariates,” but rather to “use them correctly when they satisfy the mean equivalence/no effect criteria.” This perspective bridges the gap between theory and practice, offering a principled alternative to the status quo that can deliver substantial efficiency gains.

In our empirical study, we applied this selection procedure conservatively. Even under this cautious approach, we observed large and systematic reductions in estimator variance across multiple experiments. These findings suggest that further gains may be achievable by identifying additional informative post-treatment covariates, either through richer domain expertise or data-driven methods such as machine learning.

The framework also offers practical advantages in terms of computational efficiency and scalability. Because it relies on a small number of carefully vetted covariates and avoids the need to train complex predictive models for each experiment, the method fits naturally within large-scale A/B testing platforms where thousands of experiments run in parallel. By providing a unified approach that is theoretically grounded, easy to implement, and compatible with existing experimentation pipelines, our work contributes both conceptual clarity and immediate practical value to the design and analysis of online experiments.

% Acknowledgments---Will not appear in anonymized version
\acks{This work was primarily conducted during the first author’s summer internship at Etsy. The
first author gratefully acknowledges the support from the Two Sigma PhD Fellowship. The
authors would like to thank Peng Ding for his insightful comments, and thank Stephane
Shao, Ami Wulf, Li Zhang, and Kelly McManus at Etsy for their valuable discussions. The
authors also would like to thank Julie Beckley and Kevin Gaan at Etsy for their support
throughout the internship program.}

\bibliography{ref}

\appendix

\newpage

\section{Technical Appendices and Supplementary Material}

\subsection{Proof of Proposition~\ref{prop:1}}

\begin{proof}
    Let $\overline{Y} = n^{-1} \sum_{i=1}^n Y_i$ and $\overline{Z} = n^{-1} \sum_{i=1}^n Z_i$ denote the sample means of $Y$ and $Z$, respectively. Let $\mathbb{P}$ and $\mathbb{P}_n$ represent the population distribution and the empirical distribution of $X$. For any measurable function $f$, we have $\mathbb{P} f = \E[f(X)]$ and $\mathbb{P}_n f = n^{-1} \sum_{i=1}^n f(X_i)$.

By the definition of $\hat\gamma$, we have
\begin{align*}
    \hat\gamma = &\Big(\frac{1}{n} \sum_{i=1}^n (Z_i - \overline Z) (Z_i - \overline Z)^\top \Big)^{-1} \\
    &\Big(\frac{1}{n} \sum_{i=1}
    ^n (Z_i - \overline Z) (Y_i - \hat f(X_i) - \overline Y + \mathbb{P}_n \hat f) \Big)\\
    =&\Big(\frac{1}{n} \sum_{i=1}^n (Z_i - \overline Z) (Z_i - \overline Z)^\top \Big)^{-1} \\
    &\Big(\frac{1}{n} \sum_{i=1}
    ^n (Z_i - \overline Z) (Y_i - f(X_i) - \overline Y + \mathbb{P}_n f)\Big) \\
    +& \Big(\frac{1}{n} \sum_{i=1}^n (Z_i - \overline Z) (Z_i - \overline Z)^\top \Big)^{-1} \\
    &\Big(\frac{1}{n} \sum_{i=1}
    ^n (Z_i - \overline Z) (f(X_i) - \hat f(X_i) - \mathbb{P}_n f + \mathbb{P}_n \hat f)\Big),
\end{align*} 
where the first term corresponds to the true coefficient $\gamma$, and the second term represents the estimation error due to using $\hat{f}$ instead of $f$.

By the properties of least squares estimation, we have
\begin{align*}
&\Big(\frac{1}{n} \sum_{i=1}^n (Z_i - \overline Z) (Z_i - \overline Z)^\top \Big)^{-1} \\
&\Big(\frac{1}{n} \sum_{i=1}
    ^n (Z_i - \overline Z) (Y_i - f(X_i) - \overline Y + \mathbb{P}_n f)\Big) \\
    =& \gamma + O_\P(n^{-1/2}).
\end{align*}

Next, we analyze the second term. By the law of large number, the inverse of the covariance matrix $(n^{-1} \sum_{i=1}^n (Z_i - \overline Z) (Z_i - \overline Z)^\top)^{-1} = O_\P(1)$.

Note that by the Cauchy-Schwarz inequality,
\begin{align*}
&\Big\lVert\frac{1}{n} \sum_{i=1}
    ^n (Z_i - \overline Z) (f(X_i) - \hat f(X_i))\Big\rVert_2^2 \\
    \le&  \Big(\frac{1}{n} \sum_{i=1}^n \lVert Z_i - \overline Z \rVert_2^2\Big) \Big(\frac{1}{n} \sum_{i=1}^n (f(X_i) - \hat f(X_i))^2\Big).
\end{align*}

We have
\[
    \frac{1}{n} \sum_{i=1}^n \lVert Z_i - \overline Z \rVert_2^2 = {\rm Tr}\Big(n^{-1} \sum_{i=1}^n (Z_i - \overline Z) (Z_i - \overline Z)^\top\Big) = O_\P(1),
\]
and since $\mathbb{P}_n$ converges weakly to $\mathbb{P}$,
\begin{align*}
    &\frac{1}{n} \sum_{i=1}^n (f(X_i) - \hat f(X_i))^2 \\
    =& \lVert \hat f - f \rVert_2^2 + (\mathbb{P}_n - \mathbb{P}) [(\hat f - f)^2] \\
    =& \lVert \hat f - f \rVert_2^2 + o_\P(1). 
\end{align*}

Therefore,
\begin{align*}
&\Big(\frac{1}{n} \sum_{i=1}^n (Z_i - \overline Z) (Z_i - \overline Z)^\top \Big)^{-1} \\
&\Big(\frac{1}{n} \sum_{i=1}
    ^n (Z_i - \overline Z) (f(X_i) - \hat f(X_i))\Big) \\
    =& O_\P(\lVert \hat f - f \rVert_2) + o_\P(1).
\end{align*}

Additionally, we note that
\[
    \frac{1}{n} \sum_{i=1}
    ^n (Z_i - \overline Z) (\mathbb{P}_n \hat f - \mathbb{P}_n f) = 0.
\]

Combining all the above results, we conclude that 
\begin{align*} 
    \hat\gamma - \gamma = O_\P(\lVert \hat f - f \rVert_2) + o_\P(1). 
\end{align*}
\end{proof}

\subsection{Proof of Proposition~\ref{prop:2}}

\begin{proof}
    We have
\begin{align*}
    &\P(p_j \le \alpha, \forall j \in \cS_1) = 1 - \P(p_j > \alpha, \exists j \in \cS_1) \\
    \ge& 1 - \sum_{j \in \cS_1} \P(p_j > \alpha).
\end{align*}

Since the tests are consistent by the assumption, for any $j \in \cS_1$, we have $\lim_{n \to \infty} \P(p_j > \alpha) = 0$. As the set $\mathcal{S}_1$ is finite (with cardinality at most $m$), it follows that
\[
    \lim_{n \to \infty}\P(p_j \le \alpha, \forall j \in \cS_1) = 1.
\]

This completes the proof.
\end{proof}

\subsection{Proof of Proposition~\ref{prop:3}}

\begin{proof}
    For any test with significance level $\alpha$, we have
    \begin{align*}
        &\limsup_{n \to \infty } \P(p_j \le \alpha, \exists j \in \cS_0) \le \limsup_{n \to \infty } \sum_{j \in \cS_0} \P(p_j \le \alpha) \\
        \le& \lvert \cS_0 \rvert \alpha = m\alpha.
    \end{align*}

Similarly, from the consistency of the tests, we have
\[
    \limsup_{n \to \infty} \P(p_j > \alpha, \exists j \in \cS_1) = 0.
\]

Combining these results, we obtain for any $\alpha \in (0,1)$,
\begin{align*}
    &\liminf_{n \to \infty} \P(\{j:p_j > \alpha\} = \cS_0) \\
    \ge& 1 - \limsup_{n \to \infty} \P(p_j \le \alpha, \exists j \in \cS_0) - \limsup_{n \to \infty} \P(p_j > \alpha, \exists j \in \cS_1) \\
    \ge& 1-m\alpha.
\end{align*}

Finally, by letting $\alpha \to 0$, we have
\begin{align*}
    \lim_{\alpha \to 0} \lim_{n \to \infty}\P(\{j:p_j > \alpha\} = \cS_0) = 1.
\end{align*}

This completes the proof.
\end{proof}

\end{document}